\documentclass[a4paper,11pt]{article}
\usepackage{jheppub}
\usepackage{mathrsfs}
\usepackage{amsfonts}
\usepackage{setspace}
\usepackage{cellspace}
\usepackage{amsmath,bm}
\usepackage[colorlinks=true,linkcolor=blue]{hyperref}
\usepackage{xcolor}
\usepackage{epsfig}
\usepackage{slashed}
\usepackage{caption}
\usepackage{hhline,multirow,tabularx}  % for nicer tables
\usepackage{dcolumn}    % align table columns on decimal point
\usepackage{url}        % for URL addresses
\usepackage{braket}

\title{A note on the 2D NLSM free energy }
\author[a]{Yizhuang Liu}

\affiliation[a]{Institute of Theoretical Physics,
Jagiellonian University, 30-348 Kraków, Poland}

\emailAdd{yizhuang.liu@uj.edu.pl}

\abstract {This note contains the perturbative computation of the 2D non-linear sigma model (NLSM) energy-density in a chemical potential $h$, at the fourth order in the coupling constant expansion. The result is in fully agreement with the $h\rightarrow \infty$ asymptotics extracted from the thermodynamical Bethe ansatz (TBA).   }
\date{\today}

\begin{document}
\maketitle
\flushbottom

\section{NLSM perturbative free energy}
In this note, we compute the $\alpha^4$ order perturbative correction to the 2D NLSM energy-density in a chemical potential, and test against the TBA. This is the first order at which the $\zeta_3$, as well as the NNLO contribution in the large-$N$ expansion appears. We found perfect agreement with the TBA result, after converting to the same coupling constant scheme. The four-loop $\overline{\rm MS}$-scheme beta function computed in~\cite{Bernreuther:1986js} is used for the conversion. 

\subsection{The setup of the computation}
\begin{figure}[htbp]
    \centering
    \includegraphics[height=9.0cm]{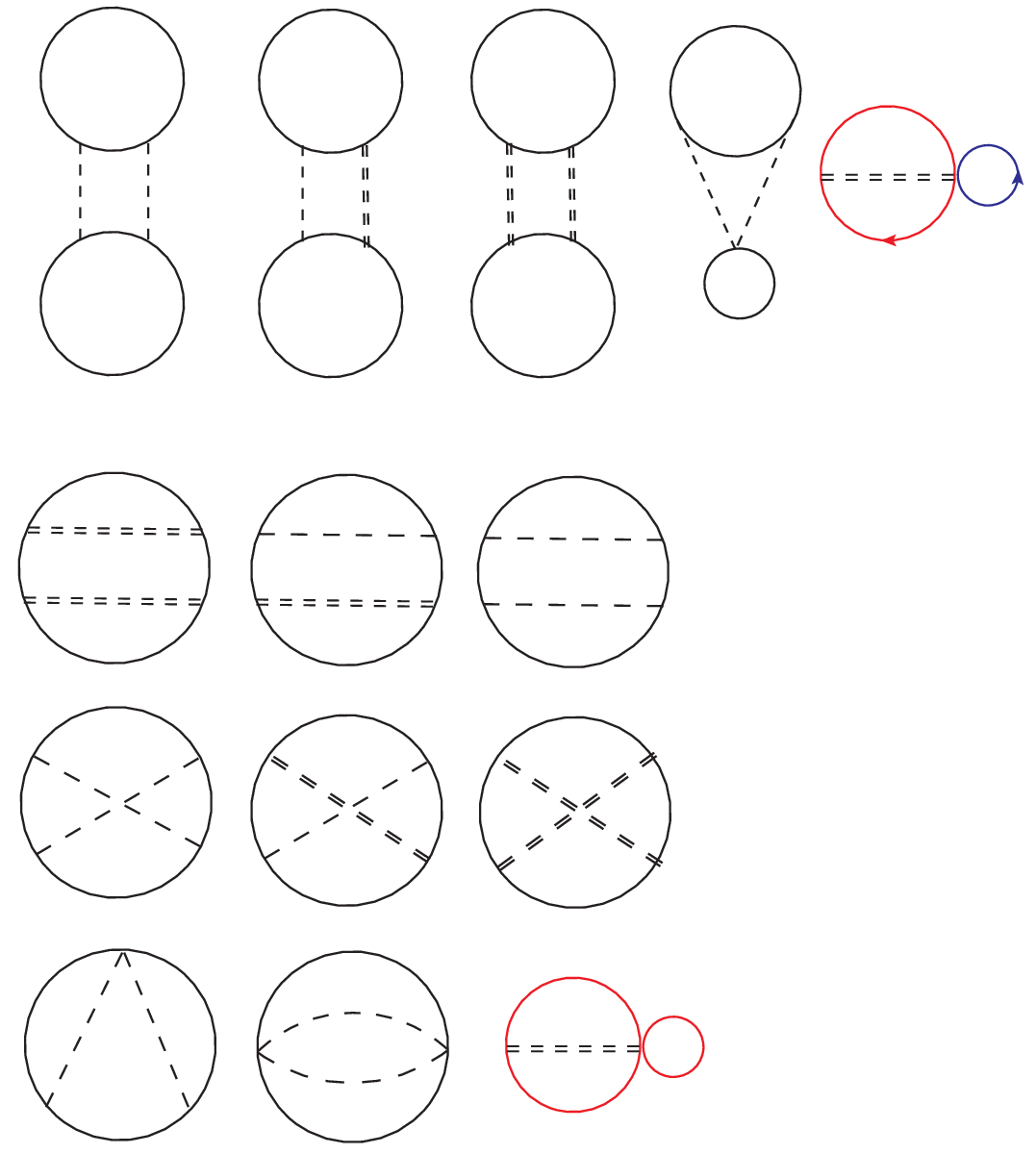}
    \caption{The diagrams for the perturbative computation. The first line are NLO in the large $N$, while the rest are NNLO. The last two diagrams in the first line are just the $\lambda^2$ order results multiplied by the $\langle \Phi^2\rangle$ tadpole. The single dashed line represents the $\theta\theta$ propagator, while the double dashed line represents the ``$k^2$-propagator''.}
    \label{fig:diagram}
\end{figure}

The Euclidean action density is~\cite{Hasenfratz:1990zz,Hasenfratz:1990ab,Bajnok:2008it} 
\begin{align}
{\cal S}=\frac{1}{2\lambda^2}
\bigg((\vec{\nabla} S_\perp )^2+(\vec{\nabla} S_\parallel)^2+2ih(\vec{S}_\parallel\times \partial_0 \vec{S}_\parallel)_3+h^2(S_\perp^2-1)\bigg) \ , 
\end{align}
where $S_\perp^a=(S^3,S^4,...S^N)$ and $\vec{S}_\parallel=(S^1,S^2)$ are the $N$-components of the scalar field living on the unit sphere, and $(\vec{a}_{\parallel}\times \vec{b}_\parallel)_3=a_1b_2-a_2b_1$. The $h$ is the chemical potential for the rotation in the $12$ plane. It is convenient to introduce a two-dimensional unit vector $\vec{n}^2=1$ for the longitudinal components $\vec{S}_{\parallel}=\sqrt{1-S_\perp^2}\vec{n}$,  and rewrite the action as
\begin{align}
{\cal S}=\frac{1}{2\lambda^2}\bigg((\nabla S_\perp)^2+h^2(S_\perp^2-1)+\frac{(\nabla S^2_\perp)^2}{1-S_\perp^2}+(1-S_\perp^2)((\nabla \vec{n})^2+2ih(\vec{n}\times \partial_0\vec{n})_3)\bigg) \ .
\end{align}
To perform the perturbative expansion, we introduce 
\begin{align}
S_\perp^a=\lambda \Phi^a,  \  \vec{n}=\left(\cos \lambda\theta, \sin \lambda \theta\right) \ . 
\end{align}
If we set to zero the total derivative term 
\begin{align}
S_{-1}=\frac{ih}{\lambda}\int d^2x \partial_0 \theta=0 \ ,  
\end{align}
which should be legal in perturbation theory, we can write the action as 
\begin{align}
&{\cal S}=-\frac{1}{2\lambda^2}+\frac{1}{2}\left((\nabla \Phi)^2+h^2\Phi^2\right)+\frac{1}{2}(\nabla \theta)^2-i\lambda h\Phi^2 \partial_0\theta-\frac{\lambda^2}{2} \Phi^2 (\nabla \theta)^2 \nonumber \\ 
&+\frac{\lambda^2 (\Phi \nabla \Phi)^2}{2(1-\lambda^2 \Phi^2)} \ . \label{eq:action}
\end{align}
The above allows the perturbative expansion order by order in $\lambda=\lambda_0$, using the following propagators
\begin{align}
&\langle \Phi^a(x)\Phi^b(y)\rangle=\delta^{ab}\int \frac{d^dp}{(2\pi)^d}\frac{e^{ip\cdot (x-y)}}{p^2+h^2} \  , \\ 
&\langle \theta(x)\theta(y)\rangle=\int \frac{d^dp}{(2\pi)^d}\frac{e^{ip\cdot (x-y)}}{p^2} \ . 
\end{align}
Here we are interested in the $\lambda^4$ order correction to the free energy
\begin{align}
-{\cal F}(h)=\frac{1}V{}\ln \int D\Phi D\theta e^{-\int d^d x {\cal S}[\Phi,\theta]} \ . 
\end{align}
They are given by the connected vacuum diagrams.
At the order $\lambda^4$, there are the following terms required
\begin{align}
&A_4=3h^4\times\frac{(i)^4}{4! V} \int d^d x_1d^dx_2d^dx_3d^dx_4 \bigg\langle \Phi^2(x_1)\Phi^2(x_2)\Phi^2(x_3)\Phi^2(x_4)\bigg\rangle_c \nonumber \\ 
& \times \int\frac{d^d k_1}{(2\pi)^d}\frac{e^{ik_1(x_1-x_2)}(k_1^0)^2}{k_1^2} \int\frac{d^d k_2}{(2\pi)^d}\frac{e^{ik_2(x_3-x_4)}(k_2^0)^2}{k_2^2} \ , 
\end{align}
and 
\begin{align}
&A_3=3h^2\times\frac{(-1)^3(-i h)^2}{3!V}\int d^d x_1d^dx_2d^dx_3\bigg\langle \Phi^2(x_1)\partial_0\theta(x_1) \Phi^2(x_2) \partial_0\theta(x_2) \nonumber \\ 
&\bigg(-\frac{\Phi^2(x_3)}{2}(\nabla \theta (x_3))^2+\frac{(\nabla \Phi ^2)^2(x_3)}{8}\bigg)\bigg\rangle_c \ , 
\end{align}
and 
\begin{align}
&A_2=\frac{(-1)^2}{2!V}\int d^dx_1 d^d x_2\langle S_{2}(x_1)S_2(x_2)\rangle_c \ , \\
&S_2=-\frac{1}{2}\Phi^2(\nabla \theta)^2+\frac{1}{8}( \nabla \Phi^2)^2 \ , 
\end{align}
also 
\begin{align}
A_1=-\frac{1}{8V}\int d^dx  \langle \Phi^2( \nabla \Phi^2)^2\rangle_c \ . 
\end{align}
The contractions can be  represented in Feynman diagrams. We use the dashed single line to represent the $\theta$ propagator, solid single line to represent the $\Phi$ propagator, and dashed double line to represent the ``effective propagator'' $k^2$, also called an ``interaction line'' in~\cite{David:1980rr}, due to the $\Phi^{2m}\nabla^2 \Phi^{2n}$ vertices in the last term of Eq.~(\ref{eq:action}). There are $14$ distinct diagrams. Two of them are the $\langle \Phi^2\rangle $ tadpole multiplying the $\lambda^2$-order result. Together with another three two-ring diagrams, there are five diagrams that are NLO in the large $N$ expansion. The remaining diagrams are next-to-next-leading order in the large $N$. The diagrams are shown in Fig.~\ref{fig:diagram}.

\subsection{The various contributions}
Here we list all the contributions. First, we use $d=2-2\epsilon$ throughout the computation. Since there is only one scale $h$, we can set $h=1$ in the computation, and at the end multiply all the terms by the common factor $h^{3d-4}$. We use the following notation for the massive and massless denominators
\begin{align}
D_k=k^2+h^2; \  d_k=k^2 \ . 
\end{align}
It is convenient to introduce 
\begin{align}
I_0=\int \frac{[dk]}{D_k}=\frac{\Gamma(1-\frac{d}{2})}{(4\pi)^{\frac{d}{2}}} \ ,
\end{align}
for the massive $\langle \Phi^2 \rangle$ tadpole. The following identities are frequently used.
\begin{align}
\int \frac{[dk]}{D_k^2}=-\frac{d-2}{2}I_0 \ , \ \int \frac{[dk] k^2}{D_k}=-I_0 \ . 
\end{align}
We then introduce the 
\begin{align}
\Sigma(p^2)=\int\frac{[dk]}{D_kD_{p+k}} \ , 
\end{align}
for the one-loop massive bubble diagram, which is a familiar quantity in the bubble-chain computation. A related quantity is 
\begin{align}
\Sigma_{(1)}(p^2)=\int\frac{[dk]}{D_k^2D_{p+k}}  \ .  \label{eq:defs1}
\end{align}
There is a useful integration-by-part identity
\begin{align} \label{eq:IBPs1}
-(p^2+4)\Sigma_{(1)}(p^2)=(d-3)\Sigma(p^2)+\frac{d-2}{2}I_0 \ . 
\end{align}
We also need the following 
\begin{align}
\Pi^{ij}(p)=\int [dk]\frac{k^ik^j}{D_{p+k}d_k}=A(p^2)p^ip^j+B(p^2)\delta^{ij} \ .  \label{eq:defAB}
\end{align}
Since $k^kk^i=d_k$, we have the relation 
\begin{align}
A(p^2)p^2+dB(p^2)=I_0=\int \frac{[dp]}{D_p} \ . 
\end{align}
Then, we also need 
\begin{align}
\int [dk]\frac{k^i}{D_{p+k}d_k}=\tilde A(p^2)p^i \ \label{eq:deftildeA}. 
\end{align}
The $\Sigma$ and the $A$, $\tilde A$, $B$ can all be expressed through the hypergeometric function $\,_2F_1$. We also need the following 
\begin{align}
C(d)=\int [dk] \Sigma^2(k^2) \ . \label{eq:defC}
\end{align}
At $d=2$, it is convergent and can be evaluated in terms of the famous combination $7\zeta_3$ 
\begin{align}
C(d)|_{d\rightarrow 2} \rightarrow \frac{7\zeta_3}{(4\pi)^3} \ , 
\end{align}
that appears in many places.

\subsubsection{The two-ring diagrams}
The two-ring diagrams are NLO in the large-$N$ expansion and computed in the literature already~\cite{Marino:2021six}. They are shown in the first line of Fig.~\ref{fig:diagram}.

First, there are the two diagrams which are just the $I_0$ multiplying the $\lambda^2$ order results:
\begin{align}
I_{14}=\frac{1}{2}I_0^3 \ , I_{8}=-\frac{1}{d}I_0^3 \ . 
\end{align}
They are in the $A_1$ and $A_3$ groups. Then, we have the two-ring with two $\theta$ propagators
\begin{align}
I_1=\int[dk]\Sigma^2(k^2) \frac{(k^0)^4}{(k^2)^2} \ . 
\end{align}
Using the Lorentz invariance, we can reduce the above as  
\begin{align}
&I_1=\frac{3}{d^2+2d}\int [dk] \Sigma^2(k^2)=\frac{3}{d^2+2d}C(d) \ , 
\end{align}
where $C(d)$ is defined in Eq.~(\ref{eq:defC}). 
We then need another two-ring with two $k^2$ ``propagators''
\begin{align}
I_{10}=\int [dp] (p^2)^2\Sigma^2(p^2) \equiv I_a \ . 
\end{align}
The two-ring with one $k^2$ and another $\theta$ propagator is 
\begin{align}
I_{5}=\int[dk]\Sigma^2(k^2)(k^2)\frac{k_0^2}{k^2}=\frac{1}{d}\int[dk] \Sigma^2(k^2)k^2 \equiv \frac{1}{d}I_b \ . \label{eq:defIb}
\end{align}
The full combination with the appropriate symmetry factors is then
\begin{align}
&I_{14}+I_{8}+I_1+\frac{1}{2}I_{5}+\frac{I_{10}}{16} \nonumber \\ 
&=I_{14}+I_{8}+\frac{1}{4}\int[dk]\left(\frac{\Sigma^2(k^2)}{2}\right)^2\left(\frac{4k_0^2 h^2}{k^2}+k^2\right)^2 \ . 
\end{align}
The specific combination in the last term also appears at high perturbative-orders at the NLO in the large-$N$~\cite{Marino:2021six}. The summation over the $O(N)$ indices leads to $(N-2)^2$, which is omitted here.
\subsubsection{The planar ladders}
We then introduce the diagrams that are NNLO in the large $N$. There is an overall factor $(N-2)$ 
which we omit.

First, there are three planar ladders shown in the second line of Fig.~\ref{fig:diagram}. The first one is with two $\theta$-propagators
\begin{align}
I_2=\int \frac{[dk_1][dk_2[dp]}{D_pD_{p+k_1+k_2}D_{p+k_1}^2}\frac{(k_1^0)^2(k_2^0)^2}{k_1^2k_2^2} \ .
\end{align}
Using the $\Pi^{ij}$ defined before in Eq.~(\ref{eq:defAB}), it can be expressed as
\begin{align}
&I_2=\int \frac{[dp]}{D_p^2} \left(A(p^2)p^0p^0+B(p^2)\right)^2=\int\frac{[dp]}{D_p^2} \bigg(\frac{3p^4 A^2}{d^2+2d}+\frac{2 p^2AB}{d}+B^2\bigg) \nonumber \\ 
&=\frac{2d-2}{d^2(d+2)}\int\frac{[dp]}{D_p^2}(p^2A)^2+\frac{I_0^2}{d^2}\int\frac{[dp]}{D_p^2} \ . 
\end{align}
The second term contains  UV divergences, while the first term is convergent. For $d=2$, we need
\begin{align}
A(p^2)p^2 \rightarrow \frac{1}{4\pi } \bigg(-1+\left(1+\frac{1}{p^2}\right)\ln\left(p^2+1\right) \bigg) \ . 
\end{align}
Then, the integral at $d=2$ can be computed as 
\begin{align}
\int\frac{dp}{D_p^2}(p^2 A)^2 \rightarrow \frac{1}{(4\pi)^3 } \ . 
\end{align}
Another similar integral with $D_p$ in the denominator is much harder and will be encountered in another diagram. 

The second planar ladder is the mixed one 
\begin{align}
I_6=\int [dp][dk_1][dk]\frac{(k_1^0)^2 k^2}{D_{p-k_1}k_1^2  D_p^2D_{p-k}}=\frac{I_0^2}{d}\int[dp][dk_1]\frac{(D_p-2)}{D_p^2}=\frac{(d-1)I_0^3}{d} \ . 
\end{align}
The third one is with two $k^2$ propagators and is also simple
\begin{align}
&I_{11}=\int [dp][dk_1][dk_2]\frac{k_1^2k_2^2}{D_{p+k_1+k_2}D_{p+k_1}^2D_p}=I_0\int [dk_1][dp]\frac{k_1^2((p+k_1)^2-1)}{D^2_{p+k_1}D_p} \nonumber \\ 
&=I_0^2\int [dp] \frac{(p^2-1)^2}{D_p^2}=-2dI_0^3 \ . 
\end{align}
In total, the combination of the planar ladders with the symmetry factors is
\begin{align}
4I_2+2I_6+\frac{1}{4}I_{11} \ . 
\end{align}
The integrals in the category happens to be simpler than others.

\subsubsection{The crossed ladders}
There are three crossed ladders,  shown in the third line of the Fig.~\ref{fig:diagram}.

First, the one with two $\theta$ propagators
\begin{align}
I_3=\int [dk_1][dk_2][dp] \frac{(k_1^0)^2(k_2^0)^2}{k_1^2k_2^2}\frac{1}{D_pD_{p+k_1}D_{p+k_2}D_{p+k_1+k_2}} \ . 
\end{align}
This can be evaluated using the following decomposition under average of the type $f(k_1^2,k_2^2,k_1\cdot k_2)$
\begin{align}
&k_1^ik_1^j k_2^lk_2^m \nonumber \\ 
&\rightarrow \delta^{ij}\delta^{lm}\left(\frac{(d+1)k_1^2k_2^2-2 (k_1\cdot k_2)^2}{d \left(d^2+d-2\right)}\right)+(\delta^{il}\delta^{jm}+\delta^{im}\delta^{jl})\left(\frac{d(k_1\cdot k_2)^2 -k_1^2k_2^2}{d \left(d^2+d-2\right)}\right) \ , 
\end{align}
which gives the following 
\begin{align}
(k_1^0)^2(k_2^0)^2 \rightarrow \frac{k_1^2k_2^2+2 (k_1\cdot k_2)^2}{d^2+2 d} \ . 
\end{align}
The leads to 
\begin{align}
&I_3=\frac{1}{d^2+2d}\int [dk] \Sigma^2(k^2) \nonumber \\ 
&+\frac{1}{d^2+2d}\int[dk_1[dk_2][dp] \frac{2k_1\cdot k_2 (k_1\cdot k_2)}{k_1^2k_2^2D_pD_{p+k_1}D_{p+k_2}D_{p+k_1+k_2}} \ . 
\end{align}
The last line can be reduced using 
\begin{align}
2k_1\cdot k_2=D_{p+k_1+k_2}+D_p-D_{p+k_1}-D_{p+k_2} \ . 
\end{align}
The four resulting terms are equal and lead to
\begin{align}
\frac{4}{d^2+2d}\int[dp][dk_1][dk_2]\frac{k_1\cdot k_2}{D_pk_1^2D_{p+k_1}k_2^2D_{p+k_2}}=\frac{4}{d^2+2d}\int[dp]\frac{p^2\tilde A^2(p^2)}{D_p} \ .
\end{align}
Here we have used the $\tilde A$ function defined in Eq.~(\ref{eq:deftildeA}). As such, we have 
\begin{align}
I_3=\frac{1}{d^2+2d}C(d)+\frac{4}{d^2+2d}\int[dp]\frac{p^2\tilde A^2(p^2)}{D_p}  \ . 
\end{align}
At $d=2$, we have 
\begin{align}
\tilde A(p^2)=\frac{\ln(1+p^2)}{4\pi p^2} \ . 
\end{align}
Then, 
\begin{align}
&\frac{4}{d^2+2d}\int[dp]\frac{p^2\tilde A^2(p^2)}{D_p} \rightarrow \frac{4}{d^2+2d}\frac{2\zeta_3}{(4\pi)^{\frac{3d}{2}}} \ , \\
&I_3 \rightarrow \frac{15\zeta_3}{8(4\pi)^3} \ . 
\end{align}
We also have the crossed ladder with one $\theta$ and one $k^2$ propagator
\begin{align}
&I_{7}=\frac{1}{d}\int [dl_1][dl_2][dk] \frac{(l_1-l_2)^2}{D_{l_1}D_{l_1+k}D_{l_2}D_{l_2+k}}  \nonumber \\
&=\frac{1}{d}\left[2I_0^3-\int[dk] \Sigma^2(k^2)\left(2+\frac{k^2}{2}\right)\right] \equiv \frac{I_b}{d} \ . \label{eq:I12identity} 
\end{align}
Notice that $I_b$ is defined in Eq.~(\ref{eq:defIb}), and the last equality is based on an identity of $I_b$ to be introduced latter.

Finally, we need the one with two $k^2$ propagators
\begin{align}
I_{12}=\int [dp][dk_1][dk_2]\frac{k_1^2k_2^2}{D_pD_{p+k1}D_{p+k_2}D_{p+k_1+k_2}} \ . 
\end{align}
We can write it in the following form as 
\begin{align}
I_{12}=\int [dl_1][dl_2][dk] \frac{k^2(l_1-l_2)^2}{D_{l_1}D_{l_1+k} D_{l_2}D_{l_2+k}} \ . 
\end{align}
The above can be computed using 
\begin{align}
(l_1-l_2)^2k^2 \rightarrow k^2(D_{l_1}+D_{l_2}-2m^2)-\frac{(k^2)^2}{2} \ , 
\end{align}
as 
\begin{align}
I_{12}=-4I_0^3-\int[dk]\Sigma^2(k^2)\left(\frac{(k^2)^2}{2}+2k^2\right) \  . 
\end{align}
All together, the crossed ladders contribute to
\begin{align}
2I_3+I_{7}+\frac{1}{8}I_{12} \ , 
\end{align}
where the symmetry factors are taken into account. 

\subsubsection{The last line of Fig.~\ref{fig:diagram}}
Finally, we introduce the remaining diagrams shown in the last line of Fig.~\ref{fig:diagram}.  The simplest is the last diagram resulting from one contraction in the $A_1$ group. It is given by 
\begin{align}
I_{13}=\int [dk_1][dk_2][dp]\frac{k_1\cdot k_2}{D_pD_{p+k_1}D_{p+k_2}}=-I_0^3 \ . 
\end{align}
The most interesting two diagrams are due to the $\Phi^2 (\nabla \theta)^2$ vertices. The contraction of two such vertices gives 
\begin{align}
I_9=\int [dp]\Sigma(p) \int [dk]\frac{(k\cdot(p+k))^2}{k^2(p+k)^2} \ .
\end{align}
It can be computed in closed form as
\begin{align}
I_9=\frac{1}{(4\pi)^{\frac{3d}{2}}}\frac{\pi ^{3/2} 4^{d-1}  \Gamma \left(2-\frac{3 d}{2}\right) \Gamma (2-d)}{\sin \frac{\pi d }{2}\Gamma \left(\frac{3}{2}-d\right)} \ . 
\end{align}
Finally, the contraction of one $\Phi^2 (\nabla \theta)^2$ and another two $-ih\partial_0 \theta \Phi^2$ gives 
\begin{align}
I_4=\int [dk_1][dk_2][dp]\frac{k_1^0k_2^0 k_1\cdot k_2}{D_p D_{p+k_1}D_{p+k_2}k_1^2k_2^2} \ . 
\end{align}
It can be computed using the same method as before
\begin{align}
I_4=\frac{d-1}{d^2}\int\frac{[dp] (p^2A)^2}{D_p}+\frac{I_0^3}{d^2} \ ,  \label{eq:defI4}
\end{align}
where the scalar function $A$ is defined in Eq.~(\ref{eq:defAB}). There is still an UV divergence in the first term, which should be extracted. In total, including the sign and the symmetry factors, the last line of Fig.~\ref{fig:diagram} contributes to 
\begin{align}
-I_{13}+\frac{I_9}{2}-4I_4  \ . 
\end{align}
We have enumerated all the contributions.

\subsubsection{Summary of the contributions}
According to the numbers of interaction vertices, we have the following combinations, in which the factors of $N-2$ are omitted for simplicity 
\begin{align}
&A_4=\frac{3\times 4\times 2}{4!}(I_1+4I_2+2I_3) \  , \\
&A_3=\frac{3}{3!} \left(-\frac{2\times 4\times2}{2}I_4+\frac{4 \times 2}{8} (I_{5}+4I_6+2I_{7})\right)+{\color{red} I_{8}} \ , \\
&A_2=\frac{1}{2!}\bigg(\frac{2\times2}{4}I_9+\frac{4\times 2(I_{10}+4I_{11}+2I_{12})}{8^2}\bigg) \ , \\
&A_1=-\frac{4\times 2}{8}I_{13}+{\color{red}I_{14}} \ . 
\end{align}
The above explains our labeling of the integrals. The red terms are the two diagram that are $I_0$ multiplying the $\lambda^2$-order results. The multiplicative renormalizablity would not be true without them. In fact, at the NLO in the large-$N$, such terms can be effectively absorbed into the following ``dressed-coupling''
\begin{align}
\lambda_0^2\rightarrow \frac{\lambda_0^2}{1-\lambda_0^2I_0} \ , 
\end{align}
which appeared before in~\cite{Marino:2021six}.

\subsection{Three divergent integrals }
To obtain the results, it remains to evaluate three divergent integrals with the scalar functions $(p^2A)^2$ and $\Sigma^2(p^2)$. Values for the convergent integrals at $d=2$ are already provided before.     
\subsubsection{Divergent integrals with $\Sigma^2(p^2)$ }
Here we consider the following two integrals
\begin{align}
I_a=I_{10}=\int [dp] (p^2)^2\Sigma^2(p^2) \  , \ I_b=\int [dp](p^2)\Sigma^2(p^2) \  .
\end{align}
The integral $I_a$ is quadratic divergent, while $I_b$ is logarithmicaly divergent. In the literature, there is a method~\cite{Marino:2021six} of computing these integrals with an arbitrary power of the massive bubble $\Sigma$. Here with only two of them, we can reduce them to $I_0^3$ plus the convergent integral $C(d)$ in Eq.~(\ref{eq:defC}) using the following trick. 

First, since the dimension of $I_a$ is $3d-4$, taking the $h^2$ derivatives we found
\begin{align}
\left(\frac{3d-4}{2}\right)I_a=-4\int [dp] p^2 \Sigma(p^2) p^2\Sigma_{(1)}(p^2) \ , 
\end{align}
where $\Sigma_{(1)}$ is defined in Eq.~(\ref{eq:defs1}). Using the IBP identity Eq.~(\ref{eq:IBPs1}), we found 
\begin{align}
&\frac{3d-4}{2}I_a=4(d-3)I_b+4\frac{d-2}{2}I_0\int [dp] p^2 \Sigma(p^2)+16\int [dp]p^2\Sigma(p^2)\Sigma_{(1)}(p^2)  \nonumber \\ 
&=-2dI_b+4(2-d)I_0^3 \ , 
\end{align}
where we have used the fact that the last term in the first line is again an $h^2$ derivative.
We can apply the same trick to $I_b$, which gives the simple expression
\begin{align}
I_b=\frac{4}{3}\bigg(I_0^3-\int [dp]\Sigma^2(p^2)\bigg) \equiv \frac{4}{3}\left(I_0^3-C(d)\right) \ .
\end{align}
Since $C(d)$ is finite at $d=2$, all the singularities are encoded by the $I_0^3$. The value of $C(d)$ at $d=2$ is sufficient for our computation.

\subsubsection{A divergent integral with $A^2$}
Finally, we consider the integral with $A^2$ in the contribution $I_4$ given by Eq.~(\ref{eq:defI4}). We adopts the method used in~\cite{Marino:2021six} to extract its $d-2$ expansion up to the constant order. Similar integrals also appeared in~\cite{Marino:2021six} for the ``$\psi$-$\eta$ bubbles''. 

For this, we note that in unit of $h=1$, we have 
\begin{align}
p^2A(p^2)=\frac{\Gamma(1+\epsilon)}{(4\pi)^{1-\epsilon}}\frac{p^2 \, _2F_1\left(1,2-2 \epsilon;3-\epsilon;\frac{p^2}{p^2+1}\right)}{(2-\epsilon) \left(p^2+1\right)} \ . 
\end{align}
We now change the variable to $z=\frac{p^2}{p^2+1}$, leading to
\begin{align}
&\int\frac{[dp] (p^2A)^2}{D_p}\nonumber \\ 
&=\frac{\Gamma^2(1+\epsilon)}{(4\pi)^{3-3\epsilon}\Gamma(1-\epsilon)(2-\epsilon)^2}\int_0^1 dz z^{2-\epsilon}(1-z)^{-1+\epsilon} \, _2F_1^2\left(1,2-2 \epsilon;3-\epsilon;z\right) \ . 
\end{align}
The singularity of the integral can be extracted by subtracting the $z- 1$ tail uniformly in $\epsilon$:
\begin{align}
f_t(\epsilon,z)=\frac{\pi  (\epsilon-2) (\epsilon-1) }{\sin \pi \epsilon \Gamma (2-\epsilon) \Gamma (\epsilon+1)}-\frac{(1-z)^{\epsilon} \pi   \Gamma (3-\epsilon)}{\sin \pi \epsilon \Gamma (2-2 \epsilon) \Gamma (\epsilon+1)} \ . 
\end{align}
We have up to the order of ${\cal O}(1)$
\begin{align}
\int_0^1 dz f_t^2(\epsilon,z)z^{2-\epsilon}(1-z)^{-1+\epsilon}=\frac{4}{3 \epsilon^3}-\frac{8}{3 \epsilon^2}+\frac{\frac{1}{3}+\frac{2 \pi ^2}{9}}{\epsilon}+\left(\frac{8 \zeta_3}{3}-\frac{4 \pi ^2}{9}-6\right) \ . 
\end{align}
On the other hand, the integral of the regular part $\,_2F_1^2-f_t^2$ is finite in the $\epsilon \rightarrow 0$ limit and can be evaluated in closed forms. Combining the two parts, we found the crucial expansion
\begin{align}
&\int_0^1 dz z^{2-\epsilon}(1-z)^{-1+\epsilon} \, _2F_1^2\left(1,2-2 \epsilon;3-\epsilon;z\right) \nonumber \\ 
&=\frac{4}{3 \epsilon^3}-\frac{8}{3 \epsilon^2}+\frac{\frac{1}{3}+\frac{2 \pi ^2}{9}}{\epsilon}+\frac{32 \zeta_3}{3}-\frac{4 \pi ^2}{9}+1+{\cal O}(\epsilon) \ . 
\end{align}
This method applies to higher powers of the $\,_2F_1$ as well~\cite{Marino:2021six}.

\subsection{The result in the $\overline{\rm MS}$ scheme}

Combining all the terms, we obtain the following bare results up to the constant order
\begin{align}
&-\frac{{\cal F}_{\lambda^4}(h^2)}{h^2}=(N-2)^2f(L,\epsilon)+(N-2)g(L,\epsilon) \ ,  \  L=\ln \frac{h^2}{\mu^2}  \ . \\
&f(L,\epsilon)=-\frac{1}{192 \pi ^3 \epsilon^2}+\frac{3 L-2}{192 \pi ^3 \epsilon}+\frac{63 \zeta_3-2 \left(6 L (3 L-4)+\pi ^2+20\right)}{1536 \pi ^3}\ , \\
&g(L,\epsilon)=-\frac{1}{96 \pi ^3 \epsilon}-\frac{-8 L+9 \zeta_3+2}{256 \pi ^3} \ . 
\end{align}
Notice that all the $\frac{1}{\epsilon^3}$ poles cancel at the end. Here, the expansion is performed together with the following renormalization scale
\begin{align}
\mu_0^2=\frac{e^{\gamma_E}\mu^2}{4\pi} \ , 
\end{align}
 in the $\overline{\rm MS}$ scheme. The bare coupling is given by (notice that $g$ is the square of the $\lambda$)
\begin{align}
\lambda_0^2= \mu_0^{2\epsilon} gZ_g^{-1} \ , 
\end{align}
where $Z_g^{-1}$ is the coupling renormalization constant consisting purely of $\frac{1}{\epsilon}$ poles. To extract $Z_g$ and check the multiplicative renormalizability, we need to add the results from lower orders. We have~\cite{Bajnok:2008it} 
\begin{align}
&-\frac{ {\cal F}_{\lambda^{-2}}}{h^2}=\frac{1}{2} \ . \\
&-\frac{ {\cal F}_{0}}{h^2}=(N-2)\frac{e^{\epsilon (\gamma_E -L)} \Gamma (\epsilon-1)}{8 \pi } \ , \\
&-\frac{ {\cal F}_{\lambda^2}}{h^2}=(N-2)\frac{ e^{2 \epsilon (\gamma_E -L)} \Gamma (\epsilon-1)\Gamma(\epsilon+1)}{32 \pi ^2} \ . 
\end{align}
Given all above, it is then possible to check that all the divergences can be absorbed to the following $Z_g$ which is $L$-independent
\begin{align}
Z_g=1+\frac{\alpha}{2  \epsilon}+\frac{\alpha^2 }{4 \epsilon}+ \left(\frac{\Delta (1+4 \Delta)}{24 \epsilon}-\frac{\Delta}{24 \epsilon^2}\right) \alpha^3 \ ,  \  \alpha=\frac{(N-2)g}{2\pi} \ . 
\end{align}
Moreover, the $\beta$ function extracted from the $Z_g^{-1}$ reproduces the known $\beta_2$ in the literature. Moreover, in terms of the renormalized $\alpha$ and the expansion parameter 
\begin{align}
 \Delta=\frac{1}{N-2} \ ,   
\end{align}
we have the following renormalized value
\begin{align}
&-\frac{ {\cal F}_{\alpha^2}}{h^2}=\frac{ \left(-12 L^2+24 L+63 \zeta_3-28\right)}{384 \pi ^2}+\Delta\frac{ (48 L-54 \zeta_3-12)}{384 \pi ^2} \ ,   \\
&-\frac{ {\cal F}_\alpha}{h^2}=\frac{2 L-1 }{16 \pi }\ ,  \ -\frac{ {\cal F}_0}{h^2}=\frac{L-1}{8 \pi } \frac{1}{\Delta}\  ,    \  -\frac{{\cal F}_{{\alpha^{-1}}}}{h^2}=\frac{1}{4\pi \Delta} \ . 
\end{align}
In the above, we have $\alpha=\alpha(\mu)$. Since ${\cal F}$ is RG invariant, we can set $\mu=h$ and then $L=0$. The leads to the following expansion with $\alpha=\alpha(h)$
\begin{align}
-\frac{4\pi  {\cal F}(h)}{h^2}=\frac{1}{\Delta \alpha(h)}-\frac{1}{2 \Delta}-\frac{\alpha(h)}{4}+\frac{63\zeta_3-28-\Delta(54\zeta_3+12)}{96 }\alpha(h)^2 +{\cal O}(\alpha^3)\ . \label{eq: MSbar}
\end{align}
The result is quite compact. The NNLO contribution in the large-$N$ expansion only starts at the $\alpha^2$ order. 

\subsection{Scheme conversion and comparing with TBA}
To compare with the results in the literature, we need to convert the result to the energy density, and change the coupling constant scheme to the ``minimal scheme''.  It is convenient to convert the coupling constant scheme first. For the $\alpha$ in the $\overline{\rm MS}$ scheme, we have 
\begin{align}
\mu \frac{d \alpha}{d\mu}=\beta(\alpha)=-\alpha^2-\Delta \alpha^3-\frac{\Delta(1+4\Delta)}{4} \alpha^4-\beta_3\alpha^5+....  \   \ , 
\end{align}
where the four-loop $\beta$ function value is computed in~\cite{Bernreuther:1986js} 
\begin{align}
\beta_3=-\frac{\Delta}{12} (18 \Delta(\Delta-1)  \zeta_3-6 \Delta (\Delta+3)+1) \ . 
\end{align}
The $\Lambda$ parameter in the $\overline{\rm MS}$ scheme is defined as 
\begin{align}
\Lambda=\lim_{\mu \rightarrow \infty}  \frac{\mu e^{-\frac{1}{\alpha(\mu)}}}{(\alpha(\mu))^{\Delta}} \ . 
\end{align}
Given the same $\Lambda$ parameter, we can define another $\hat \alpha$ in the ``minimal scheme''~\cite{Bajnok:2008it}
\begin{align}
\frac{1}{\hat \alpha(\mu)}+\Delta \ln \hat \alpha(\mu)=\ln \frac{\mu}{\Lambda} \ . \label{eq:hatalpha}
\end{align}
The $\alpha$ allows a perturbative expansion in terms of $\hat \alpha$ 
\begin{align}
\alpha=\hat \alpha+\frac{\Delta}{4}\hat \alpha^3+\frac{\beta_3-\Delta^3}{2}\hat \alpha^4+{\cal O}(\hat \alpha^5) \ . 
\end{align}
Since the free energy starts from the order $\alpha^{-1}$, to reach the $\hat \alpha^2$ order, we need the $\hat \alpha^4$-order term in the scheme conversion, therefore the four-loop $\beta_3$. The scheme conversion generates terms at the order $\Delta^3$ in the $\hat \alpha^2$ order free-energy. In the diagrammatic computation, such terms would be encountered only at the four-loop level. After converting the scheme, in terms of $\hat \alpha=\hat \alpha(h)$ the free energy reads
\begin{align}
&-\frac{{\cal F}(h)}{h^2}=\frac{1}{4\pi\Delta}\left(\frac{1}{\hat \alpha}-\frac{1}{2}-\frac{\Delta}{2}\hat \alpha+\eta_4 \hat \alpha^2+....\right) \ , \\ 
&\eta_4=\frac{\Delta}{32} \bigg(21\zeta_3-8-(28+42\zeta_3)\Delta+(8+24\zeta_3)\Delta^2\bigg) \ . 
\end{align}
Comparing with the $\overline{\rm MS}$ scheme expansion Eq.~(\ref{eq: MSbar}), the expansion at the $\hat \alpha^2$ order contains more terms. The $\overline{\rm MS}$ scheme absorbs much more information into the coupling constant than the minimal scheme, but still not too many to generate renormalons in its $\beta$ function. It is not unreasonable to expect, that this popular scheme works better than the minimal scheme, at scales that are not very low, but also not very asymptotic. But after penetrating the dark clouds, after the sky becomes blue and transparent, the star in the UV fixed point would no longer be screened, and will shine forever. 

Given the expansion in $\hat \alpha$, we can convert the free energy to the energy density
\begin{align}
{\cal E}(\rho)={\cal F}(h)-{\cal F}(0)+\rho h \ ,   \  \rho=- {\cal F}'(h) \ . 
\end{align}
Since ${\cal F}(0)=0$ in perturbation theory, in terms of $h$ one has 
\begin{align}
&{\cal E}(h)=\frac{h^2}{4\pi \Delta} \bigg(\frac{1}{\hat \alpha}+\frac{1}{2}+\frac{\Delta}{2}\hat \alpha+\tilde \eta_4\hat \alpha^2+...\bigg) \ , \\
&\tilde \eta_4=\frac{\Delta}{32}\bigg(21 \zeta_3+8+(4-42 \zeta _3)\Delta+(24 \zeta _3+8)\Delta^2\bigg) \ . 
\end{align}
On the other hand, using the RGE of the $\hat \alpha$, we found that 
\begin{align}
\rho=\frac{h}{4\pi \Delta}\bigg(\frac{2}{\hat \alpha}+\frac{\Delta \left(21\zeta_3-(12+42\zeta_3)\Delta+(8+24\zeta_3)\Delta^2\right)}{16} \hat \alpha^2+ \ . ..\bigg) \ . \label{eq:density}
\end{align}
We can introduce another coupling in terms of the density
\begin{align}
\tilde \alpha(\rho)+(\Delta-1)\ln \tilde \alpha(\rho) =\ln \frac{2\pi \Delta\rho}{\Lambda} \ . 
\end{align}
Combining Eq.~(\ref{eq:density}) and the Eq.~(\ref{eq:hatalpha}), we found that $\hat \alpha=\tilde \alpha+{\cal O}(\tilde \alpha^5)$. As such, combining the above, one has
\begin{align}
&{\cal E}(\rho)=\rho^2\pi \Delta\left(\tilde \alpha+\frac{\tilde \alpha^2}{2}+\frac{\Delta}{2}\tilde \alpha^3+\chi_4 \Delta \tilde \alpha^4+ ...\right) \ , \\
&\chi_4=-\frac{21\zeta_3-8{\color{red}-(28+42\zeta_3)\Delta}+(8+24\zeta_3)\Delta^2}{32} \ . 
\end{align}
The above is in fully agreement with the TBA results~\cite{Volin:2009wr}. In particular, this computation also confirms the consistency of the $\beta_3$ value computed in the literature. In fact, the whole computation is for the $-(28+42\zeta_3)\Delta$ term: the $\Delta^2$ term is due to the scheme conversion, while the $21\zeta_3-8$ is NLO in the large $N$. For the NLO in large-$N$ part, the perturbation theory has been tested with TBA up to very high orders~\cite{Marino:2021six}.  

\acknowledgments
The author thanks Zoltán Bajnok for notification of the reference~\cite{Bajnok:2008it}.

\bibliographystyle{apsrev4-1}
\bibliography{bibliography}
\appendix

\end{document}